\documentclass[12pt, a4paper]{article}

\pdfoutput=1

\usepackage[english]{babel} 
\usepackage{microtype} 
\usepackage{amssymb}
\usepackage{amsmath,amsfonts,amsthm,mathrsfs} 
\usepackage[svgnames]{xcolor} 
\usepackage[hang, small, labelfont=bf, up, textfont=it]{caption} 
\usepackage{booktabs} 
\usepackage{lastpage} 
\usepackage{graphicx} 
\usepackage{enumitem} 
\setlist{noitemsep} 
\usepackage{graphicx}

\usepackage{hyperref}
\hypersetup{
    colorlinks=true,
    linkcolor=blue,
    filecolor=magenta,
    urlcolor=cyan,
}

\usepackage{geometry} 

\usepackage[T1]{fontenc} 
\usepackage[utf8]{inputenc} 

\usepackage{XCharter} 

\usepackage{titling} 

\usepackage{authblk}

\usepackage{lettrine} 
\usepackage{fix-cm}	

\usepackage{xstring} 
\usepackage{booktabs}
\usepackage{multicol}
\usepackage{svg}

\pdfoutput=1

\usepackage{filecontents}

\title{Multi-scale Convolutional Neural Networks for Inverse Problems} 

\author[1,2]{Feng Wang\footnote{\href{mailto:feng.wang@empa.ch}{feng.wang@empa.ch} }}
\author[2]{Alberto Eljarrat}
\author[2]{Johannes M\"{u}ller}
\author[1]{Trond R. Henninen}
\author[1]{Erni Rolf}
\author[2]{Christoph T. Koch}
\affil[1]{Electron Microscopy Center, Empa, Swiss Federal Laboratories for Materials Science and Technology, CH-8600 Dubendorf, Switzerland}
\affil[2]{Institut f\"{u}r Physik, IRIS Adlershof der Humboldt-Universität zu Berlin, 12489 Berlin, Germany}

\date{\today}

\begin{document}

\maketitle 

\textbf{Inverse problems exist in many domains such as phase imaging, image processing, and computer vision. These problems are often solved with application-specific algorithms, even though their nature remains the same: mapping input image(s) to output image(s). Deep convolutional neural networks have shown great potential for highly variable tasks across many image-based domains, but are usually difficult to train due to their inner high non-linearities. We propose a novel neural network architecture highlighting fast convergence as a generic solution addressing image(s)-to-image(s) inverse problems of different domains. Here we show that this approach is effective at predicting phases from direct intensity measurements, imaging objects from diffused reflections and denoising scanning transmission electron microscopy images, with just different training datasets. This opens a way to solve problems statistically through big data, in contrast to implementing explicit inversion algorithms from their mathematical formulas. Previous works have targeted much more on \textit{how} can we reconstruct rather than \textit{what} can be reconstructed. Our strategy offers a paradigm shift.\\}

\begin{figure}[ht]
\captionsetup{format=plain}
\centering
\includegraphics[width=1.0\textwidth]{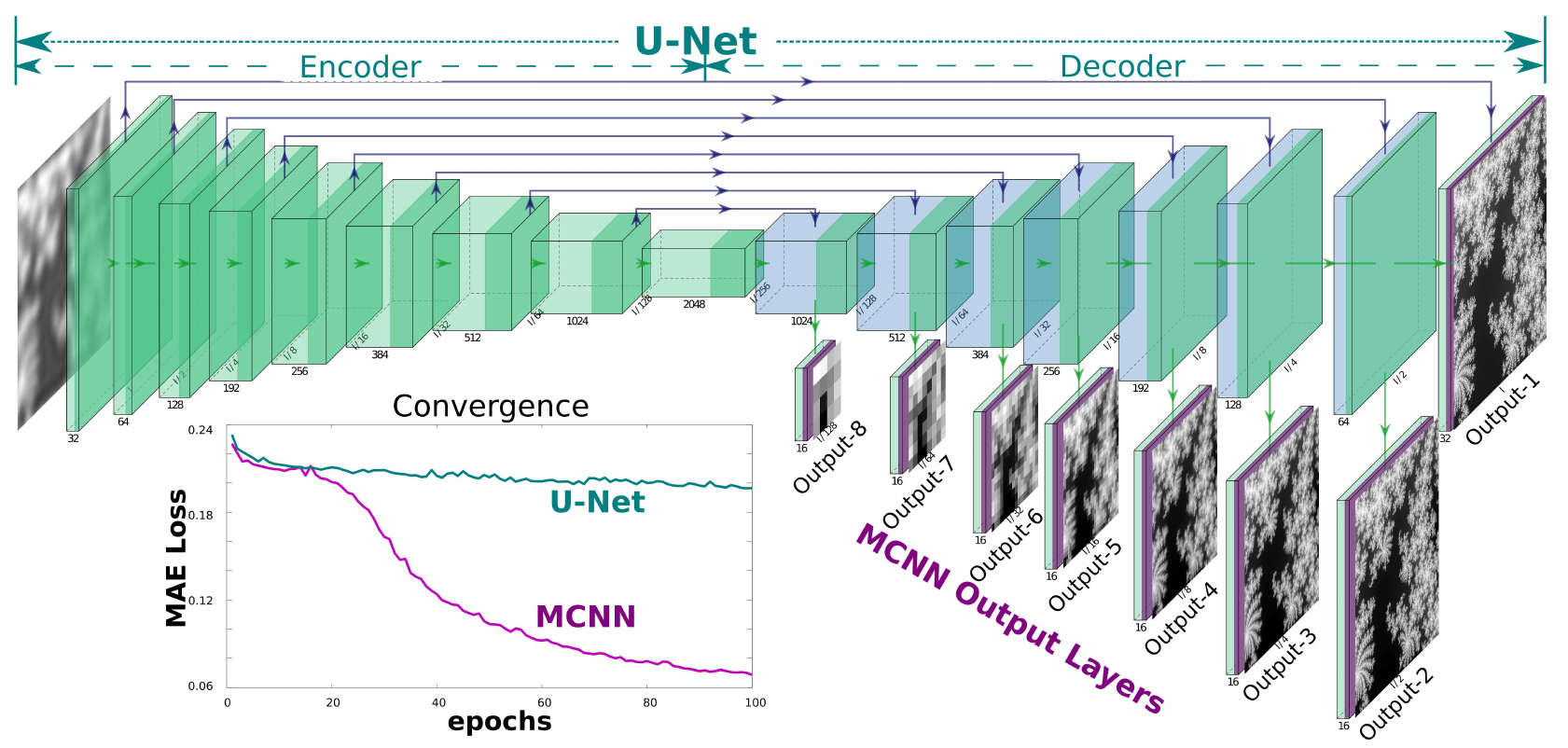}
\caption{ \textbf{By matching outputs at all frequencies, MCNN achieves better stability and faster convergence than U-Net.}
A simplified architecture of an MCNN is demonstrated. This application is designed to predict phases from a defocal image.
The network is composed of a classic U-Net (the upper part) with an additional 7 branches for multi-scale reconstruction (from Output-2 to Output-8).
With this topology, the input image is first encoded into a high dimension tensor, then is decoded into 8 images of different sizes to match different frequency components of the desired phases.
The convergence curves of the test set show the significant advantages of an MCNN over a classic U-Net: the MAE of MCNN drops quickly in 100 iterations, while the U-Net converges slowly and can get stuck at local minima.
}
\label{fig_topology}
\end{figure}

Our physical theories allow us to make predictions; given a complete description of the state of a physical system, we can make predictions of some measurements. Predicting the result of measurements is a simulation problem. On the other hand, inverse problems consist of using the actual result of some measurements to deduce the parameters characterizing the system.
And those intrinsic parameters are not obtainable using direct measurements: either because they are lost during an experiment, for example the uncertainty relation of the quantum mechanical wave equations, or perhaps because they are never measured accurately in the first place, for example, the defection in the loss of information process due to the imperfections of the imaging systems.
The simulation is explicit, and for most of the cases, is much faster than inversion, but only gives out observable details;
the inversion provides meaningful information, but is usually solved by the least-squares or probabilistic approaches, which tends to be slow and even under-determined due to the noises and its inner non-linearity. Inverse problems are some of the most important mathematical problems in science and mathematics, but most of the inverse applications are specially designed in a problem-oriented way, and there does not exist a framework as a generic solver. This means conventional inverse applications are deliberately implemented to make the best use of their domain-specific knowledge, and lack of generalization ability across different inverse applications.

Motivated by recent advances of hardware and driven by the large datasets\cite{russakovsky_imagenet_2015,krasin_openimages_2016}, deep learning\cite{lecun_deep_2015} have recently shown great potential particularly in image(s)-to-image(s) translation tasks, such as super resolution\cite{dong_image_2016}, image denoising\cite{zhang_beyond_2017} and image generation\cite{pix2pix2017}. Many inverse applications\cite{hezaveh_fast_2017,sinha_lensless_2017} have been purposed and achieved promising results, since the universal approximation theorem guarantees that neural networks can approximate arbitrary functions well\cite{hornik_multilayer_1989,cybenko_approximation_1989,lu_expressive_2017,lee_wide_2019}. However, because of the high non-linearity of such deep architectures, their performance crucially depends on proper hyper-parameter configurations. Recent advanced big models even require years to develop and tens of thousands of dollars to train\cite{yang_xlnet:_2019,radford2019language}. Although many tricks exist to tune the hyper-parameters such as the learning rate, the regulations, the design of the hidden units, the convolutional kernels and the network architectures\cite{luo_neural_2018,jinEfficientNeuralArchitecture2018}, but there is no one size fits all in favour of both the speed-up and convergence\cite{smith_disciplined_2018}.

\begin{figure}[ht]
\centering
\includegraphics[width=1.0\textwidth]{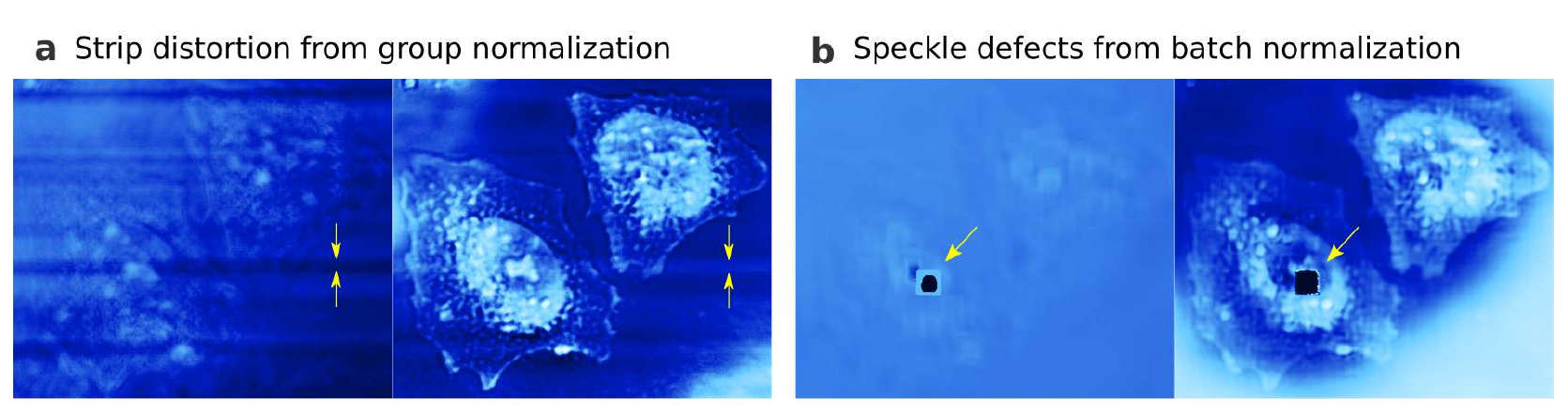}
\caption{ \textbf{Extended Figure: Distortions introduced by normalization layers.}
Inserting normalization layers before or after the activation layers will improve the convergence of the networks, but further research is desired to deal with the distortions introduced by the normalization layers.
This can be directly observed from the intermediary results when training to predict the phases and amplitudes from 8 defocal HeLa cell images.
\textbf{a}: Stripe distortion as a result of group normalization. \textbf{b}: Speckle defects from batch normalization.
}
\label{fig_distortion}
\end{figure}

Aware of the fact that a deep convolutional neural network (DCNN) first quickly captures the dominant low-frequency components, and then relative slowly captures high-frequency ones\cite{xu_training_2018,rahaman_spectral_2018}, and inspired by the idea of multi-grid methods\cite{stuben_multigrid_1982}, a novel multi-scale deep convolutional neural network (MCNN) has been designed.
This architecture extends the functionality of the hidden layers in the decoder of a U-Net\cite{hinton_reducing_2006,ronneberger_u-net_2015} by connecting these hidden layers to additional convolution layers to produce coarse outputs, in an attempt to match the low-frequency components, as is demonstrated in Fig.\ref{fig_topology}.
Another possible variation is mutating the architecture progressively by starting to fit only the components of low frequencies at the initial phases, then inserting new layers to match the increasingly high-frequency features during the training, as has been demonstrated in progressive generative adversarial networks (GAN)\cite{karras_progressive_2017}. This architecture speeds up the network convergence and greatly stabilizes the training process. As is shown in lower-left of Fig.\ref{fig_topology}, with identical setups, when solving an inverse Laplacian problem from a second order gradient approximation, MCNN quickly reaches a mean-absolute-error (MAE) loss around 0.07 while the conventional U-Net is difficult to train, being trapped around 0.2. Additional normalization layers can also accelerate the convergence\cite{ioffe_batch_2015,wu_group_2018}, but will very likely introduce undesired distortions (extended data Fig.\ref{fig_distortion}), and should be very carefully used. A topological setup of those coarse MCNN output branches are visualized in the lower part right of Fig.\ref{fig_topology}, in which the outputs matching the low-frequency components of different scales are shown; while the U-Net shown in the upper part directly matches the input layer to the desired output of the same resolution. The convolution layers inside each cell and the depth of the network are flexible of design but are generally restricted by the hardware. In contrast to conventional problem-oriented inverse applications, MCNN is designed intentionally to be problem neutral. This architecture aims to solve every inverse problem falling into the category of image(s)-to-image(s) conversion, without being limited to specific applications, but relying on massive dataset from fast numerical simulation or from direct measurement. This generalization capability is demonstrated by solving three different inverse problems.

\textbf{Solving phase problem}

The first problem is to retrieve phases of a propagated wave from the direct intensity measurements\cite{karle_recovering_1986}. This is one of the fundamental inverse problems in optics, neutron, atomic and electron optics, and has attracted a lot of efforts and led to the invention of numerous methods to reconstruct the missing phases from intensity measurements.
Following Dennis Gabor's proposal of the holographic schemes\cite{gabor_new_1948}, numerous methods have put more efforts extracting phases by post-processing images\cite{van_dyck_new_1993,lubkTransportIntensityPhase2013,jingshan_transport_2014,eljarratMultifocusTIEAlgorithm2018}.
These methods oftentimes only work under deliberate approximations and/or elaborate experimental configurations.
A more recent application based on deep learning\cite{rivenson_phase_2018} takes the complex wave of the back-propagated hologram intensities as inputs, rather than directly mapping the measured intensities to desired phases. In this application, a \textit{direct resolution} is demonstrated by mapping measurements to phases straightforwardly in an end(s)-to-end(s) manner.

\begin{figure}[ht]
\captionsetup{format=plain}
\centering
\includegraphics[width=1.0\textwidth]{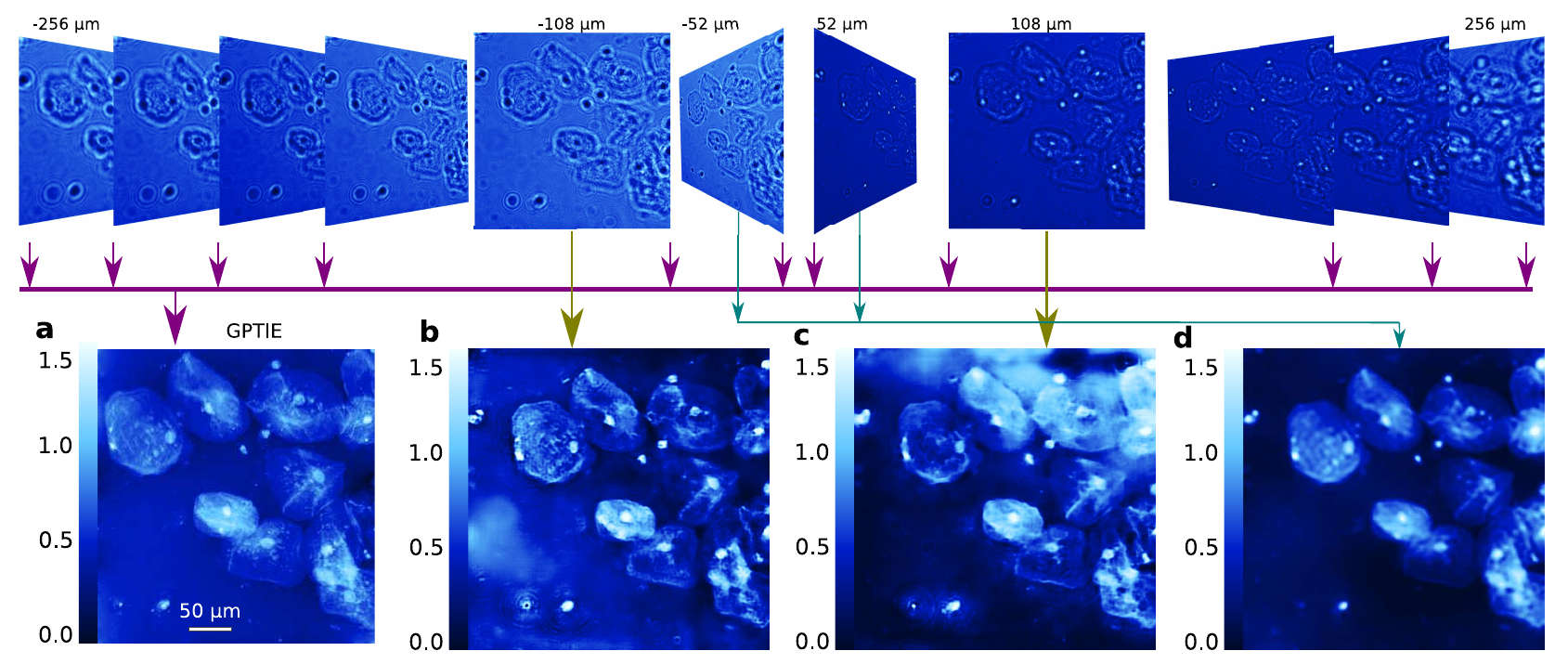}
\caption{\textbf{Phase predicted from de-focused image(s).}
From one or more de-focused images of a pure phase object made up of human cheek cells acquired equally spaced by $d_z=4\,\mu m$ from $-256\,\mu m$ to $256\,\mu m$ (only 11 shown), MCNNs are capable of phase prediction.
\textbf{a}: GPTIE reconstruction from gradients estimated using 129 images from different focal planes, with a range-of-interest of $945 \times 888$ pixels selected\cite{jingshan_transport_2014}.
\textbf{b}: Phase prediction from 1 image at a distance of $-108 \mu m$, with $1024 \times 1024$ pixels.
\textbf{c}: Prediction from 1 image at a distance of $108 \mu m$.
\textbf{d}: Prediction from 2 images at $-52 \mu m$ and $52 \mu m$ , showing good performance on par with the state-of-the-art reconstruction algorithm demonstrated in \textbf{a}.
}
\label{fig_tie}
\end{figure}

\begin{figure}[ht]
\captionsetup{format=plain}
\centering
\includegraphics[width=1.0\textwidth]{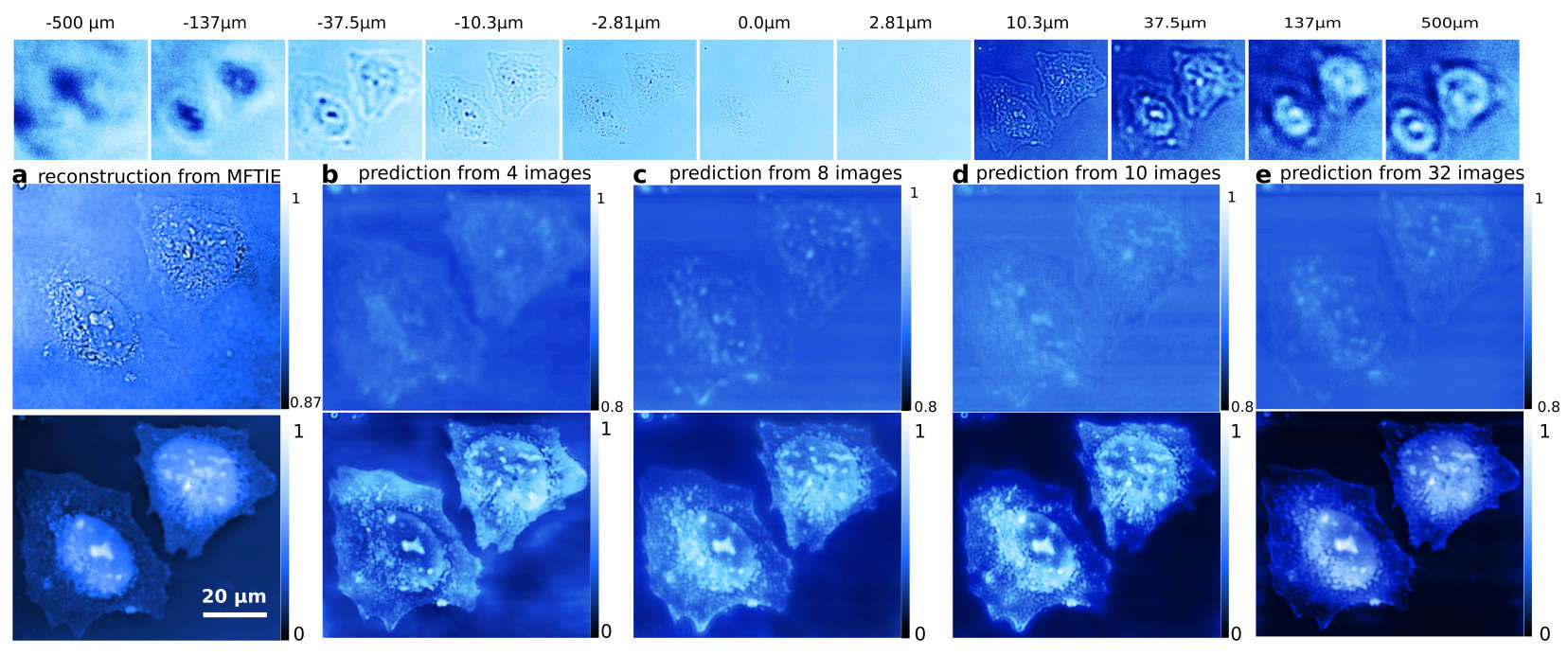}
\caption{\textbf{Amplitudes and phases predicted from de-focused images.}
First row: 51 defocal Hela cell images exponentially spaced from $-500\,\mu m$ to $500 \, \mu m$ (only 11 are shown).
\textbf{a}: Phases (top) and amplitudes (bottom) reconstructed by MFTIE from estimated gradients using all 51 images\cite{eljarratMultifocusTIEAlgorithm2018}. A range-of-interest of $470 \times 520$ pixels is selected;
\textbf{b}: MCNN prediction using 4 images taken from $-1.30 \, \mu m$ to $1.30 \,\mu m$;
\textbf{c}: Prediction using 8 image taken from $-6.13 \, \mu m$ to $6.13 \,\mu m$;
\textbf{d}: Prediction using 10 images taken from $-10.28 \, \mu m$ to $10.28 \,\mu m$;
\textbf{e}: Prediction using 32 images taken from $-48.6 \, \mu m$ to $48.6 \,\mu m$.
}
\label{fig_piah}
\end{figure}

In conventional phase retrieval schemes such as transport of intensity equation (TIE), for the sake of a good second-order gradient estimation, multiple intensities have to be measured in pairs symmetrically to the focal plane. Then from a good initial guess and with a complex non-linear optimizer, the phases can be reconstructed from a second-order, elliptic partial differential equation using the estimated gradient information.
MCNN does not suffer from such restrictions. Phases can be predicted from arbitrary intensities recorded at distance(s) of convenience, and a neural network does not need an initial guess.
Additionally, since predicting the phase from its second-order gradient is a straight forward application of MCNN, this problem has been included as a tutorial in our open source code-base.

The first phase retrieval example is presented in Fig.\ref{fig_tie}, in which three MCNNs have been trained to predict the phases from different defocal images of human cheek cells.
The first one makes use of the intensity image recorded at $-108 \, \mu m$ from the focal plane, the second one uses the image at $108 \, \mu m$, and the last one uses two images at $-52 \, \mu m$ and $52 \, \mu m$ respectively. The predicted phases are shown in \textbf{b}, \textbf{c} and \textbf{d}, which are visually similar to the result given by Gaussian-Process method\cite{jingshan_transport_2014} (GPTIE) shown in \textbf{a}.

The second example is presented in Fig.\ref{fig_piah}, in which four MCNNs have been trained to predict the amplitudes and the phases of HeLa cells.
The first one makes use of the intensities recorded in range $[-1.3 \, \mu m, 1.3 \, \mu m]$ from the focal plane, the second one uses images in range $[-6.13 \, \mu m, 6.13 \, \mu m]$, the third one uses images in range $[-10.28 \, \mu m, 10.28 \, \mu m]$ and the last one uses images in range $[-48.6 \, \mu m, 48.6 \, \mu m]$.
Comparing with the results shown in \textbf{a}, which is produced by multi-focus TIE (MFTIE)\cite{eljarratMultifocusTIEAlgorithm2018} using all the 51 images,
these applications give out, albeit from different images, very similar phases, which are shown from \textbf{b} to \textbf{e}.
The predicted amplitudes are quite blurry compared to the MFTIE result. This is an expected behaviour as the neural networks are trained by minimizing the MAE, with which the optimizer will try to minimize the averaged error, giving equal weights to all pixels despite some pixels having particularly high errors.

\textbf{Imaging objects from diffuse reflection}

The second problem is to image objects that are hidden from direct view using their indirect diffuse light reflections.
Observing objects located in inaccessible regions makes a lot of sense in the field of remote sensing, computer version, and vehicular automation. This problem has drawn significant attention in recent works\cite{klein_tracking_2016,otoole_confocal_2018,saunders_computational_2019,heide_non-line--sight_2019}, but most of these applications require complicated inversion algorithms and special experimental setups and light sources. With MCNN, the prediction of a series of two-dimensional colourful objects from non-line-of-sight (NLOS) imaging is made possible without requiring controlled or time-varying illumination, high-speed sensing and complicated inversion method from ray optics.

\begin{figure}[ht]
\captionsetup{format=plain}
\centering
\includegraphics[width=1.0\textwidth]{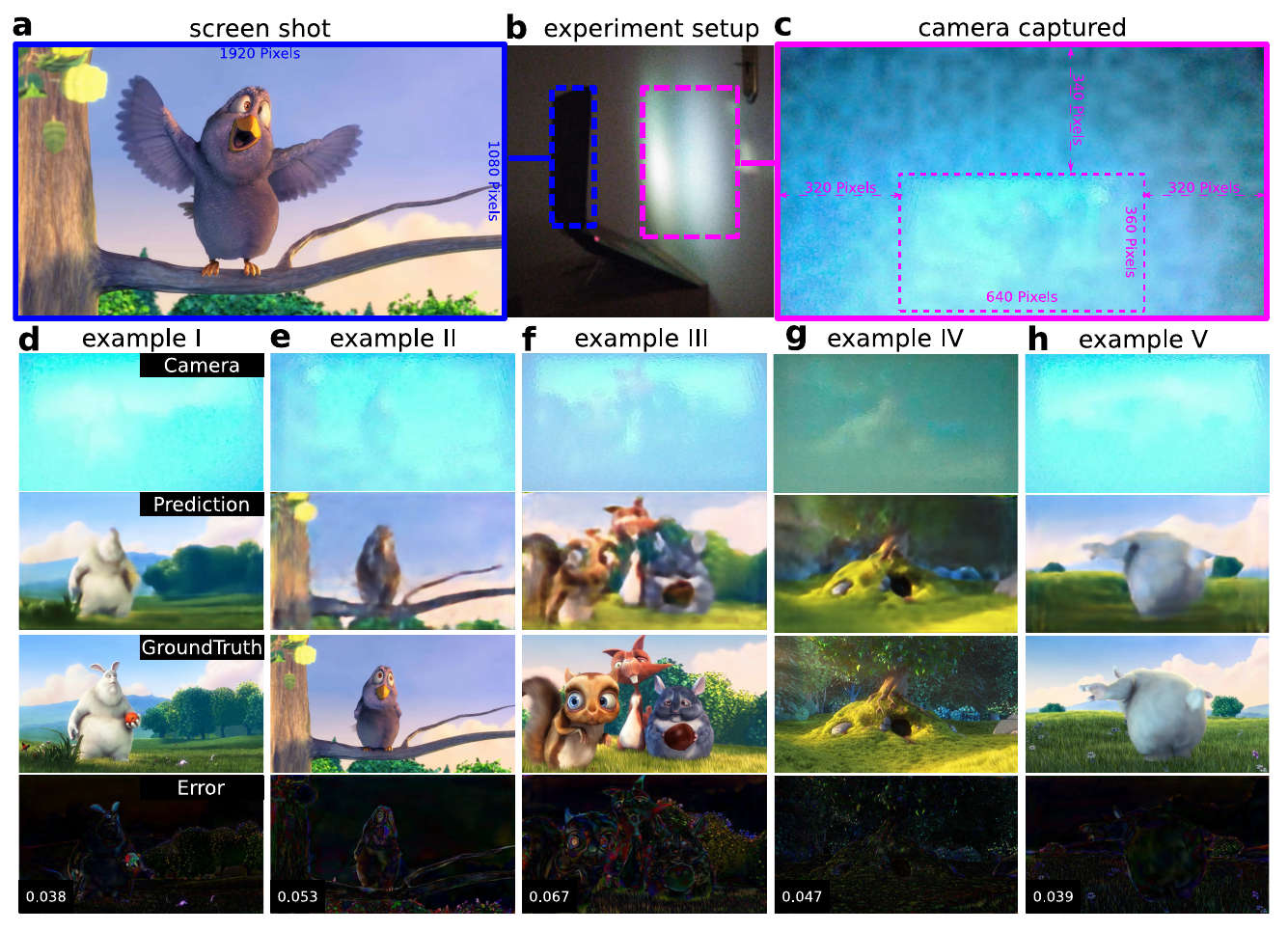}
\caption{\textbf{Imaging objects from diffuse reflections.}
This is done by training an MCNN matching the camera captured diffuse reflective images to the screenshot images.
\textbf{a}: A screenshot image. This is one of the target images matching the MCNN output.
\textbf{b}: Experimental setup. Everything is done with a laptop facing a door, no additional devices required.
\textbf{c}: A camera captured image corresponding to the screenshot shown in \textbf{a}. The cropped zone marked with a dotted rectangle is one of the input images of the MCNN.
\textbf{d}--\textbf{h}: Predictions randomly sampled from test set. The first row shows the selected range of the camera captured images (cropped and flipped), the second row shows the predictions, the third row shows the screenshots and the last row shows the absolute difference between the predictions and the screenshots, with the MAE value presented at the lower-left corner.
}
\label{fig_door_mirror}
\end{figure}


Using an ordinary laptop, with 1024 images captured by its camera and 1024 screenshots recorded from its screen, MCNN is capable of predicting objects, the screenshots, from diffusely reflected rays, the camera-captured images.
The data collection procedure is shown in \textbf{b} of Fig.\ref{fig_door_mirror}: a laptop is facing a door with a program running on it to record images from the screen and from the camera at 2 frames-per-second (fps). One pair of the captured images are shown in \textbf{a} and \textbf{c}.
The predicted images from the testing set match the ground truth well when the background is changing smoothly, as can be interpreted from the visualized absolute difference shown at the last row from \textbf{d} to \textbf{h}, where most differences are close to zero (black), especially the regions corresponding to the sky, the cloud, and the green grass.
But the high-frequency details are missing in the predicted images.
For example, in \textbf{d}, the ear of the rabbit and the red apple are lost; in \textbf{e} and \textbf{g}, the details to the eyes are lost; in \textbf{f}, all details of the animals are blurry; and in \textbf{g} and \textbf{h}, all details of the green grass are lost. More results on the test set are included in extended data video 1.


\textbf{Denoising STEM images}

\begin{figure}[ht]
\captionsetup{format=plain}
\centering
\includegraphics[width=1.0\textwidth]{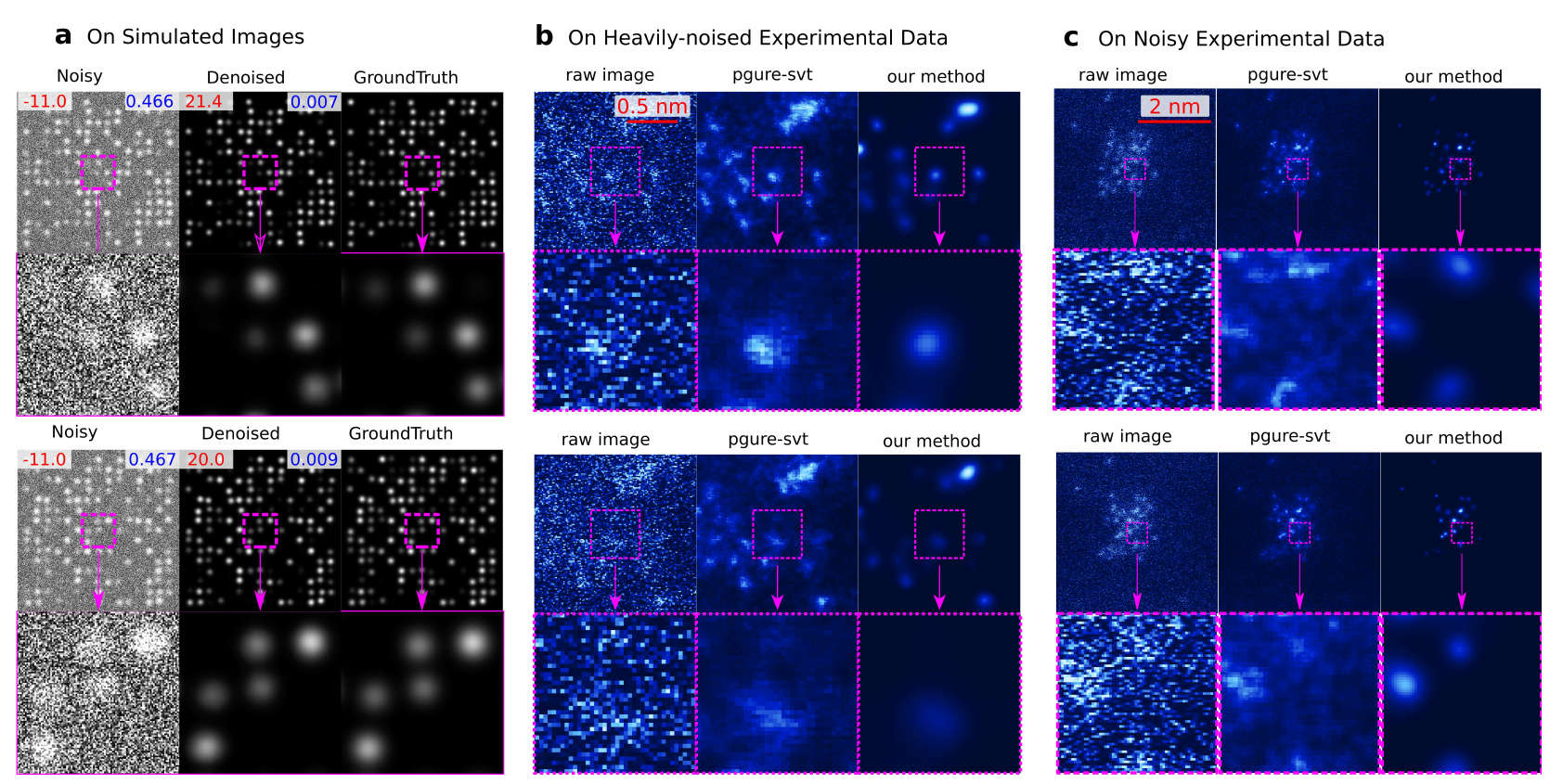}
\caption{\textbf{Denoising STEM images.}
MCNN performs well on heavily-noised datasets, as is demonstrated in \textbf{a}. The numbers on the upper-left corner are the SNRs and the upper-right are MAEs.
Also MCNN gives out clear and consistent results on consecutive aberration-corrected HAADF STEM image frames shown in the left columns, which are recorded at 150 fps \textbf{b} with 128 $\times$ 128 pixels and 15 fps \textbf{c} with 512 $\times$ 512 pixels, and are taken under electron dose in range $[10^5, 10^6] e$\r{A}$^{-2} s^{-1}$ with a FEI Titan Themis. The upper images are for the first frames, the lower images are for the second frames.
Similar denoising results produced by PGURE-SVT are shown in the middle columns, but are not as clear as MCNN results given in the right columns.
More examples of experimental data are include in extended data video 2 and 3.
}
\label{fig_denoising}
\end{figure}

The third problem is to denoise heavily-noised scanning transmission electron microscopy (STEM) images.
Modern STEM can provide sub-\r{a}ngstr\"{o}m imaging resolution\cite{erni_atomic-resolution_2009,jiang_electron_2018}, but this is limited by, e.g, the beam sensitivity of the specimen.
Lowering the electron dose results in noisy images with a poor signal-to-noise ratio (SNR) less than 0 dB, and complicated extraction of relevant specimen information.
Moreover, STEMs can produce millions of images in a few hours at a speed of 100 fps, and this amount of data can require months to process for a conventional algorithm.
It is therefore important to predict desired images from noisy observations very quickly without loss of information\cite{trond_in_prepare}.
Due to complex unknown environmental variables of specific STEM setups, real-world STEM image denoising is too complex for a single monolithic denoising algorithm.
Many conventional algorithms exist to address this problem, but they heavily depend on domain specific knowledge and/or \textit{a priori} information\cite{mevenkamp_poisson_2015,PGURESVT2016}, and their attempt at real-time denoising is difficult.
Recent variations\cite{guo_toward_2019,UlyanovVL17,zhang_beyond_2017,hasinoff_burst_2016} based on DCNN work well by directly matching a noisy input image to a clean output image, but they are constrained to high SNR images, mostly above 10.0 dB, with known noise sources, and most of them only consider a single signal independent noise with known levels.

Imaging sub-nanometre sized Platinum clusters by high-angle annular dark-field (HAADF), heavily-noised STEM images are generated at a very high speed, from 15 fps producing $512 \times 512$ pixels covering an area of 20 nm$^2$ to 150 fps with $128 \times 128$ pixels for 2.56 nm$^2$.
To visually identify dynamic behaviour, low-frequency feature extraction is essential for a successfully interpretation of atomic information.
Considering the fact that the undesired noises usually exist as bright or dark data points in the recorded images, where the high-frequency components dominates, the weights of the second layer are specially designed to mimic the behaviours of 4 different low pass filters (LPFs).
This greatly stabilized the performance of the neural network, enabling it to provide robust prediction when the noise levels varies from images to images as the experimental setup changes with time.

\begin{figure}[ht]
\centering
\includegraphics[width=1.0\textwidth]{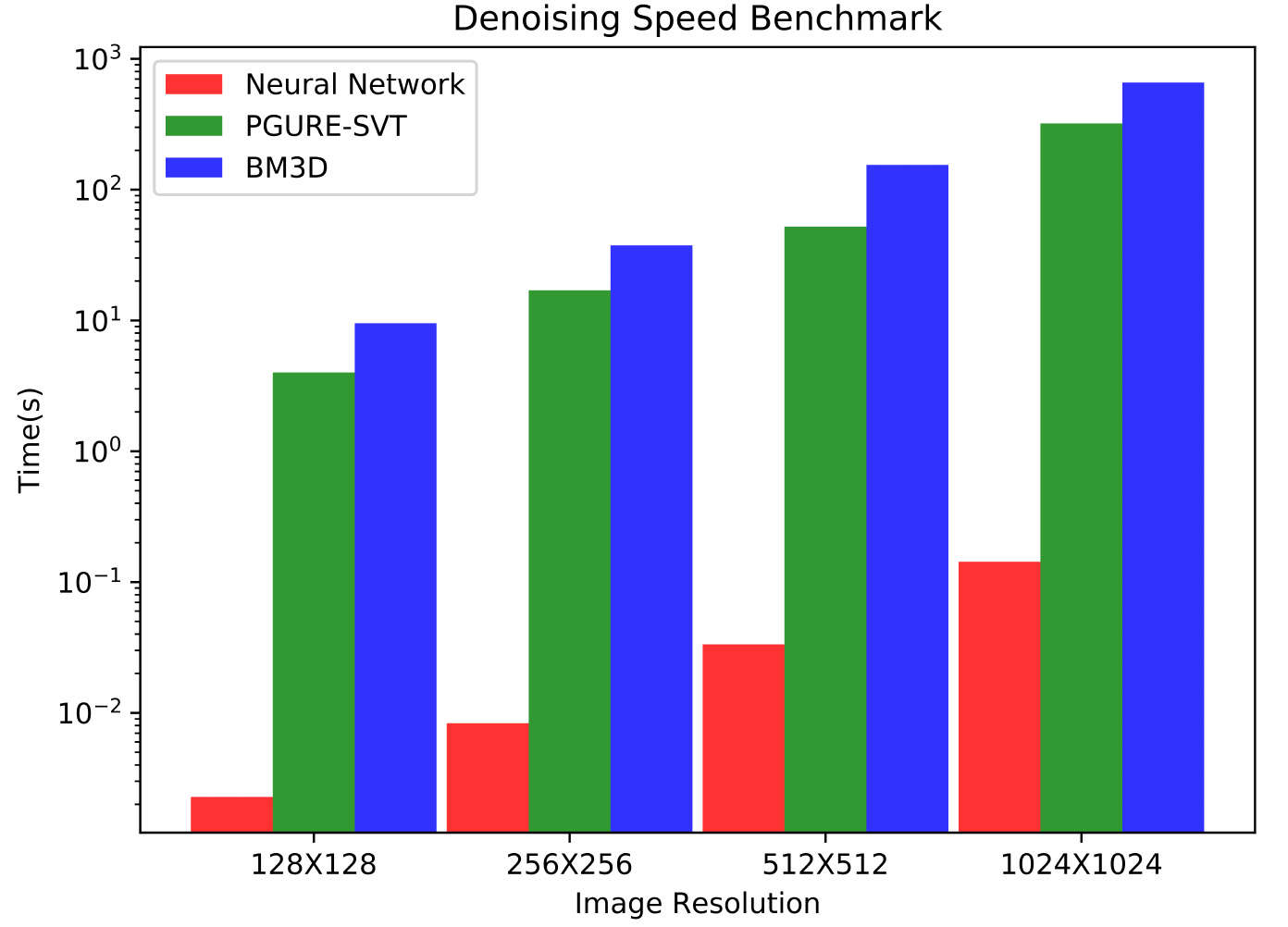}
\caption{ \textbf{Extended Figure: MCNN outperforms conventional denoising algorithms by more than three orders.}
This figure shows the average denoising time for MCNN, BM3D and PGURE-SVT methods, on experimental images from $128 \times 128$ pixels to $1024 \times 1024$ pixels.
}
\label{fig_denoising_benchmark}
\end{figure}

In this application, an MCNN is trained to predict clear results from heavily-noised STEM images recorded with less than 10 electron counts per pixel.
Testing with STEM images taken at 150 fps with $128 \times 128$ pixels, the network achieves a denoising performance up to 440 fps, more than 3 orders faster than conventional methods, taking big advantage of modern hardware acceleration (extended data Fig.\ref{fig_denoising_benchmark}).

The effectiveness of this MCNN application is validated in two ways.
First, it is tested using simulated noisy atomic images with Poison noise, Gaussian noise and white noise falling to an SNR less than -10 dB.
In this task, MCNN achieves an average SNR as high as 20 dB.
Two examples are shown in Fig.\ref{fig_denoising}\textbf{a}, in which all atoms are successfully restored, and the MAEs are less the 0.01.
Second, it is cross-validated with the state-of-the-art Poisson-Gaussian Unbiased Risk Estimator for Singular Value Thresholding (PGURE-SVT) algorithm\cite{PGURESVT2016}.
Two randomly selected consecutive frames from two videos are shown in \textbf{b} and \textbf{c}.
Visually comparing their denoised results, the neural network gives similar but much clearer atomic images on consecutive frames, demonstrating the effectiveness of MCNN, despite the predictions are only from single frames (More examples are in extended data video 2, and 3).

\textbf{Conclusion}

To conclude, an application-neutral framework has been proposed and demonstrated.
The generic capability of solving different inverse problems has been demonstrated by 3 applications in different domains.
The difficulties of a challenging application, the complex experimental setups and the complicated inverse algorithm implementation, has been alleviated with this framework.
Quick and smooth convergence is guaranteed by matching additional output layers to corresponding low-frequency features, reducing the frustration of DNN hyper-parameter tuning.
The prediction speed, taking the advantages of modern hardware and highly optimized computational frameworks, can be accelerated by multiple orders of magnitude, making real-time inverse applications feasible.
When training the MDCNN architectures presented above, weights are updated by minimizing the mean absolute value error between the prediction and the ground truth.
This error metric tends to give a loss of sharpness.
Further research, very possibly in how to integrate generative adversarial networks\cite{goodfellow_generative_2014,pix2pix2017}, is highly desirable.
In the wake of this robust, generic purposed architecture, a branch of new schemes dealing with different kinds of inverse problems are supposed to emerge, broadening the scope of inverse problem applications.
Equipped with MCNN, scientists and engineers are expected to think more about \textit{what} would be interesting to figure out, rather than \textit{how} to find out something interesting.
Intelligence resource is scarce for most of the time, when the computation resource is getting adequate, we should adapt ourselves to the trend of the paradigm shift from knowledge-driven to data-driven\cite{kitchin_big_2014}.

\bibliographystyle{naturemag}

\bibliography{reference}

\textbf{Acknowledgements}
F. W., T. H. and R. E. acknowledge founding from the European Research Council (ERC) under the European Union's Horizon 2020 program (grant agreement No. 681312).
A. E. acknowledges founding from the German research foundation (DFG), collaborative research center (SFB 951).
F. W. acknowledges the Blender Foundation, for their excellent open source animation Big Buck Bunny.

\textbf{Author contributions}
F. W. conceived the idea of MCNN.
A. E. and J. M. carried out the defocal imaging experiment.
F. W. carried out the defuse reflection experiment.
T. H. carried out the STEM experiments.
F. W. implemented all the MCNN applications.
F. W. wrote up the manuscript.
A. E. and R. E. commented on the manuscript.
R. E. and C. K. supervised the project.
The manuscript reflects the contributions of all authors.

\textbf{Competing interests} The authors declare no competing interests.

\textbf{Code and data availability} An open-source version will be soon made available at \url{https://github.com/fengwang/MCNN}. Related datasets and a minimal example can will found in this repository as well.

\textbf{Additional information}

\textbf{Extended data video 1} This video presents the predicted results from camera captured diffuse reflection images in the test set, with 256 frames displaying at 2 fps. The diffuseimages are shown on the upper-left, the ground truth is shown on the upper-right, the prediction is shown on the bottom-left, and the absolute difference is shown on the bottom-right.\newline
\textbf{Extended data video 2} This video presents 100 frames of MCNN denoised result, showing an atomic cluster that is rotating and reconstructing, displayed at 2 fps. \newline
\textbf{Extended data video 3} This video shows 100 frames of the raw STEM images, the PGURE-SVT denoised and MCNN denoised images, displayed at 2 fps. The raw images were recorded at 150 fps with $128 \times 128$ pixels.\newline
\newline

\textbf{Method}

\textbf{Dataset.} For phase retrieval applications, the training sets were simulated using Fourier optics including slight Poison noise and white noise.
The original phases and amplitudes were randomly sampled from Open Images Dataset\cite{krasin_openimages:_2017} (OID).
Millions of defocal images were simulated, but for each MCNN, only 2,048 random samples were selected to train, as no apparent performance improvement have been found when expanding the training set even to 51,200.
The exponentially spaced defocal images shown in Fig.\ref{fig_tie} come from an open access GPTIE dataset\cite{jingshan_transport_2014}.
Each image has $1024 \times 1024$ pixels with effective size $0.31 \times 0.31 \mu m^2$ per pixel.
The sample is unstained cheek cells from a human's mouth, placed on a microscope slide and sealed with a cover slide. This represents nearly a pure phase object, though there are some specks of amplitude variation.
The defocal images shown in Fig.\ref{fig_piah} come from the MFTIE dataset\cite{eljarratMultifocusTIEAlgorithm2018}. Each image has $2560 \times 2160$ pixels with effective size $0.1625 \times 0.1625 \mu m^2$ per pixel.
The sample is HeLa cancer cells which are fairly transparent.

For the diffused reflection reconstruction application, the dataset was collected using an HP OMEN 17 laptop.
With a full-screened film, \textit{Big Buck Bunny}\cite{bbb3} with 4K resolution and 60 Hz frame-rate, being played at a half-speed on a 17.3 inches LCD screen (16:9, 43.9 cm in diagonal), this laptop, mounted on a laptop stand, was positioned facing to a door at a distance about 30 cm, as is shown in Fig.\ref{fig_door_mirror} (B), and a Python script was running simultaneously, acquiring pictures (RGB) from its camera with a resolution of $1280 \times 720$ pixels and capturing screenshot images (RGB) with a resolution of $1920 \times 1280$ pixels at $2$ fps.
Multiple sources of noise exists in the acquired image-pairs coming from three main contributors.
Firstly, the sensitivity of the camera was adjusted automatically and adaptively, resulting in slightly varying brightness and contrast between frames.
This is very apparent in the first few frames.
Secondly, the laptop vibrated a lot during recording.
A powerful gaming GPU is equipped in this laptop, and a strong fan is attached to the GPU for thermal control.
This GPU was under heavy pressure when a 4K resolution film is being displayed at a screen with a refresh rate of 120 Hz, and the fan switched its working mode frequently during the acquisition time.
Lastly, there was a small but random time interval between the screen-shooting and camera capturing.
The acquisition program is coded carefully in a way ensuring both the screen images and the camera images are cached in the random access memory (RAM) before saving to the harddisk, but still, the Python script runs slowly in nature and its garbage collection behaviour is uncontrollable, and therefore the uncertainty of the timing difference is inevitable.
After $1024$ image pairs were collected, the first 256 pairs were selected for testing, and the rest $768$ pairs were selected for training by matching the images from the camera to the screenshots.
To fit the model into the GPU memory, an area of $640 \times 360$ pixels of each input images had been cropped, and the output images were scaled from $1920 \times 1280$ pixels down to $1280 \times 720$ pixels.
Besides, as a mirror reflect an image with a left/right reversal, the cropped input images were horizontally flipped. All the predicted results from the first 256 camera captured images are presented in the extended data video 1.

For the denoising application, Poison, Gaussian and white noise clipping were included in the training set generation.
Millions of images had been generated from the clear images randomly sampled from OID images and simulated annular dark-field (ADF) STEM images.
For the first training stage, the neural network was trained with half of the images from OID, and the other half from ADF images.
Including random images from OID is important, as this strategy prevents the network from predicting everything to be zero at the very beginning, because the simulated ADF images contain an average small intensities.
For the second stage, this neural network was fine-tuned on millions of ADF images with a very small learning rate.
The denoising network was tested on various experimental STEM images recorded using a probe corrected FEI Titan Themis at 300 kV, with an electron dose ranging from $10^{5} e $\r{A}$^{-2}$ to $10^{6} e $\r{A}$^{-2}$.
The samples were made by plasma sputtering Pt onto Protochips Fusion thermal chips .
The typical pixel sizes of the recorded images are 6.3-12.5 pm acquired at 15-150 fps and resolutions from $128 \times 128$ pixels to $512 \times 512$ pixels.

\textbf{Network architectures.} For the design of the network unit cells composing the MCNN shown in Fig.\ref{fig_topology}, there is a single convolutional layer with stride 2 in the encoder and a single deconvolutional layer with stride 2 in the decoder.
This is the choice of the phase-only retrieval shown in Fig.\ref{fig_tie} and the denoising application shown in Fig.\ref{fig_denoising}, with about 50 million trainable parameters inside.
For the phase-and-amplitude retrieval application shown in Fig.\ref{fig_piah} and diffuse reconstruction application shown in Fig.\ref{fig_door_mirror}, a bottleneck structure is selected in each of the unit cells, expanding the network to hundreds of layers, but reducing the trainable parameters to 20 million, to accelerate the training process and fit the model into the GPU memory. This is inspired by the recent architectures ResNeXt\cite{xie_aggregated_2017} and Xception\cite{chollet_xception:_2017}.
Also, the low-frequency output branches are reduced from 7 to 3 to fit everything into GPU memory.
For the denoising network, an extra four-channeled layer is inserted right after the input layer.
The filters of this layer are specially designed to mimic the functionality of 4 LPFs:
\begin{align*}
f_1(r,c) &= 1  \qquad & r,c \in [1,5], \\
f_2(r,c) &= 1  \qquad & r,c \in [1,7], \\
f_3(r,c) &= e^{-[{(r-8)^2+(c-8)^2}]/{\sqrt{20}}} \qquad & r,c \in [1,15], \\
f_4(r,c) &= e^{-[{(r-8)^2+(c-8)^2}]/{\sqrt{30}}} \qquad & r,c \in [1,15],
\end{align*}
in which $r$ and $c$ are the row and column index of the filters.
Furthermore, to prevent the network from predicting every pixel to be zero during the training, a GAN has been employed to adjust the back-propagated errors.
More detailed network architectures can be found within the released source code.

\textbf{Training settings.} All the networks are trained with 2 Nvidia GTX 1080 Ti GPUs using an Adam optimizer\cite{kingma_adam_2014} with its recommended parameters, except the denoising network uses a much smaller step size of 0.005 and decreases at every epoch.
The phase retrieval applications converge very quickly, and all the networks are trained only for 32 epochs in a few hours.
The diffuse reflection reconstruction application is trained for 256 epochs in half a day.
The denoising application is trained for 4096 epochs, and at each new epoch, all the 2048 examples in the training set are refreshed, ensuring only simulated STEM images exist in the later training stage. The total training time is about one week.

\end{document}